\documentclass[manuscript]{aa}

\usepackage{graphicx}
\usepackage{epstopdf}
\usepackage[intlimits,sumlimits]{amsmath}
\usepackage{deluxetable}
\usepackage{times,epsfig} 

\usepackage{amssymb}
\usepackage{mathrsfs} 
\usepackage{dirtytalk}

\usepackage{natbib}
\bibpunct{(}{)}{;}{a}{}{,} 
\usepackage{caption}

\usepackage{amsmath}
\usepackage[english]{babel}
\usepackage{longtable}
\usepackage{afterpage}
\usepackage{url}
\usepackage{hyperref}
\usepackage{xcolor}
\definecolor{dark-red}{rgb}{0.4,0.15,0.15}
\definecolor{dark-blue}{rgb}{0.15,0.15,0.4}
\definecolor{medium-blue}{rgb}{0,0,0.5}
\hypersetup{
    colorlinks, linkcolor={dark-red},
    citecolor={dark-blue}, urlcolor={medium-blue}
}

\newcommand{\beqa}{\begin{eqnarray}} 
\newcommand{\eeqa}{\end{eqnarray}}

\newcommand{\bsub}{\begin{subequations}}
\newcommand{\esub}{\end{subequations}}
\newcommand{\beal}{\begin{align}}
\newcommand{\ealn}{\end{align}}

\newcommand{\msun}{M$_{\sun}$}

\begin{document}
\title{Two stripped envelope supernovae with circumstellar interaction}
\subtitle{- but only one really shows it}

\author{J. Sollerman\inst{1} 
\and{C. Fransson}\inst{1}
\and{C. Barbarino}\inst{1}
\and{C. Fremling}\inst{2}
\and{A. Horesh}\inst{3}
\and{E. Kool}\inst{1}
\and{S. Schulze}\inst{4}  
\and{I. Sfaradi}\inst{3}
\and{S. Yang}\inst{1}
\and{E. C. Bellm}\inst{5} 
\and{R. Burruss}\inst{6} 
\and{V. Cunningham}\inst{7}
\and{K. De}\inst{2}
\and{A. J. Drake}\inst{2} 
\and{V.~Z. Golkhou}\inst{5,8}
\and{D.~A. Green}\inst{9} 
\and{M. Kasliwal }\inst{2}
\and{S. Kulkarni}\inst{2} 
\and{T. Kupfer}\inst{10}
\and{R. R. Laher}\inst{11} 
\and{F. J. Masci}\inst{11} 
\and{H. Rodriguez}\inst{6} 
\and{B. Rusholme}\inst{11} 
\and{D.~R.~A. Williams}\inst{12}
\and{L. Yan}\inst{2} 
\and{J. Zolkower}\inst{6} 
}
\institute{Department of Astronomy, The Oskar Klein Center, Stockholm University, AlbaNova, 10691 Stockholm, Sweden
\and{Cahill Center for Astrophysics, California Institute of Technology, 1200 E. California Blvd. Pasadena, CA 91125, USA}
\and{Racah Institute of Physics, The Hebrew University of Jerusalem, Jerusalem 91904, Israel}
\and{Department of Particle Physics and Astrophysics, Weizmann Institute of Science, 234 Herzl St, 76100 Rehovot, Israel}
\and{DIRAC Institute, Department of Astronomy, University of Washington, 3910 15th Avenue NE, Seattle, WA 98195, USA}
\and{Caltech Optical Observatories, California Institute of Technology, Pasadena, CA  91125, USA} 
\and{Department of Astronomy, University of Maryland, College Park, MD 20742, USA} 
\and{The eScience Institute, University of Washington, Seattle, WA 98195, USA}
\and{Astrophysics Group, Cavendish Laboratory, 19 J. J. Thomson Ave., Cambridge CB3 0HE, UK} 
\and{Kavli Institute for Theoretical Physics, University of California, Santa Barbara, CA 93106, USA} 
\and{IPAC, California Institute of Technology, 1200 E. California, Blvd, Pasadena, CA 91125, USA}
\and{Jodrell Bank Centre for Astrophysics, School of Physics and Astronomy, The University of Manchester, Manchester, M13 9PL, UK
}
}

\date{}
\abstract
    {We present observations of SN 2019tsf (ZTF19ackjszs) and SN 2019oys (ZTF19abucwzt). These two stripped envelope (SE) Type Ib supernovae (SNe) suddenly showed a (re-)brightening in their late light curves. We investigate this in the context of circumstellar material (CSM) interaction with previously ejected material, a phenomenon that is unusual among SE SNe.}
    {We use our follow-up photometry and spectroscopy for these supernovae to demonstrate the presence of CSM interaction, 
    estimate the properties of the CSM and discuss why the signals are so different for the two objects.} 
    {We present and analyse observational data, consisting of optical light curves
     and spectra. For SN\,2019oys we also have detections in radio as well as limits from UV and  X-rays.}
    {Both light curves show spectacular re-brightening after about 100 days. In the case of SN 2019tsf, the re-brightening is followed by a new period of decline, and the spectra never show signs of narrow emission lines that would indicate CSM interaction. On the contrary, SN 2019oys made a spectral makeover from a Type Ib to a spectrum clearly dominated by CSM interaction at the light
    curve brightening phase. Deep Keck spectra reveal a plethora of narrow high ionization lines, including coronal lines, and the radio observations show strong emission.}
{The rather similar light curve behaviour - with a late linear rebrightening - of these two Type Ib SE SNe indicate CSM interaction as the powering source. For SN 2019oys the evidence for a phase where the ejecta hit H-rich material, likely ejected from the progenitor star, is conspicuous. We observe strong narrow lines of H and He, but also a plethora of high ionization lines, including coronal lines, revealing shock interaction. 
Spectral simulations of SN 2019oys show two distinct density components, one with density $\ga 10
^9 \ {\rm cm}^{-3}$, dominated by somewhat broader, low ionization lines of \ion{H}{I},  \ion{He}{I}, \ion{Na}{I} and  \ion{Ca}{II}, and one with narrow, high ionization lines at a density $\sim 10
^6 \ {\rm cm}^{-3}$. The former is strongly affected by electron scattering, while the latter is unaffected by this.
The evidence for CSM interaction in SN 2019oys is corroborated by detections in radio. On the contrary, for SN 2019tsf, we find little evidence in the spectra for any CSM interaction. 
}

\authorrunning{Sollerman et al.}
\titlerunning{CSM interaction in two SE SNe.}

\maketitle

\section{Introduction}
\label{sec:intro}

Core-collapse (CC) supernovae (SNe) are explosions of massive stars ($\gtrsim8~M_\odot$) reaching the end of their stellar life-cycles. 
The variety of CC SNe 
is largely determined by the progenitor mass at the time of CC, but also by the mass-loss history leading up to the explosion. 
Hydrogen-poor CC SNe originate from massive progenitor stars that have lost most - or even all - of their H envelopes prior to explosion. 
These include Type IIb SNe (some H left), SNe Ib (no H, some He), SNe Ic (neither H nor He) as well as superluminous supernovae of Type~I (SLSNe-I). 
Collectively, SNe~IIb, Ib and Ic are called stripped-envelope (SE) SNe. 

There are few observational constraints on mass loss for very massive stars, and the processes involved are poorly understood. 
Models argue that for a star to experience enough mass loss to become a SE SN, either strong stellar winds from very massive progenitors ($\gtrsim30$~\msun,~\citealp{Groh:2013ab}), or binary interactions are needed. In the binary scenario the progenitors can be of somewhat lower mass ($\lesssim20$~\msun,~\citealp[e.g.,][]{Yoon2015}).

Evidence is emerging that a large fraction of SE SNe originate from binary systems. 
Both detailed studies of individual SNe, like the Type IIb SNe~1993J \citep{Nomoto1993,Maund:2009} and 2011dh \citep{Ergon2014,Ergon2015}, as well as sample studies (\citealp{Cano:2013aa,taddia2015,2016MNRAS.457..328L,Taddia2018,2019MNRAS.485.1559P}) indicate ejecta masses of just a few \msun. This is too low to be consistent with 
the most massive stars that lose their envelopes due to winds \citep{Groh:2013ab}. 
However, in either case, there must be ample material from the progenitor surrounding the stripped star at the time of explosion. The composition and distribution of this material contain information about the mass-loss process, as many of the binary stripping scenarios couple the phases of mass-transfer to the original binary separation \cite[e.g.,][]{Smith2014}.
The observational signatures would be evidence that the SN ejecta run into this circumstellar envelope material during some phase of the supernova evolution.
This interaction between the ejecta and the circumstellar material (CSM) can produce a significant contribution to the total luminosity \citep[e.g.,][]{ChevalierFransson2017}.

Evidence for the presence of significant CSM has been found in some SE SNe of Type IIb; late spectra 
of SN 1993J showed a broad flat-topped 
hydrogen emission line that can be explained as due to CSM interaction 
\citep[][see also \citealt{Fremling_aalrxas} for similar signatures in ZTF18aalrxas]{Matheson2000,1996ApJ...456..811H,2005ApJ...622..991F}. 

The past years of observations have also revealed cooling phases similar to those observed in the early light curves (LC) of SNe IIb among other SE SN subtypes, indicating extended material outside these otherwise compact progenitors. Examples include the Type Ic SNe iPTF15dtg \citep{Taddia2016} and iPTF14gqr \citep{De2018}, where the latter also showed so-called flash spectroscopy signatures indicative of close-by CSM \citep{GalYam2014}.
Moreover, several SLSNe-I have been found to enter into an interaction phase with H-rich CSM in the years after explosion. In these cases, broad H features developed over time 
\citep{Yan2017}. Finally, SN 2014C \citep{Milisavljecic2015}, SN 2017ens \citep{2018ApJ...867L..31C}
 and SN 2017dio \citep{2018ApJ...854L..14K} constitute three recent cases where SE SNe have spectroscopically metamorphosed into CSM interacting Type IIn supernovae, revealing the presence of external CSM at later phases.
 
In this paper we present two SE SNe that were discovered after peak in their evolution, but that both after a few months started to (re-)brighten. The extra power needed for such a light curve evolution is presumably CSM interaction simply because none of the other powering mechanisms at play at later phases are likely to display such a behaviour (see e.g., \citealt{Sollerman2019} for a discussion and assessment of some of the scenarios; magnetar, radioactivity, accretion). 

The paper is organized as follows. 
In Sect.~\ref{sec:obs} we present the observations, including optical photometry and spectroscopy but also some space-based observations and radio data. 
Section~\ref{sec:discussion} presents a discussion of the similarities and differences between the two objects and finally Sect.~\ref{sec:conclusions} presents our conclusions, 
and contains a discussion where we put our observations in context with other SNe.

\section{Observations}
\label{sec:obs}

\subsection{Detection and classification}
\label{sec:detection}

SN\,2019oys (a.k.a. ZTF19abucwzt)
was first detected on 2019 Aug 28 ($\mathrm{JD}=2458723.98$), with the Palomar Schmidt 48-inch (P48) Samuel Oschin telescope as part of the Zwicky Transient Facility (ZTF) survey \citep{2019PASP..131a8002B,GrahamM2019}. It was reported to the Transient Name Server 
(TNS\footnote{\href{https://wis-tns.weizmann.ac.il}{https://wis-tns.weizmann.ac.il}})
on Aug 29. The first detection is in $g$ band, with a host-subtracted magnitude of $19.14\pm0.12$~mag, at the J2000.0 coordinates $\alpha=07^{h}07^{m}59.26^{s}$, $\delta=+31\degr39\arcmin55.3\arcsec$. 
This transient was subsequently also reported to the TNS by several other surveys; in September by Gaia and ATLAS and in November by Pan-STARRS. 

SN 2019oys is positioned in the spiral galaxy
CGCG 146-027 NED01 that had a reported redshift of
$z = 0.0165$. Using a flat cosmology with H$_0=70$~km~s$^{-1}$~Mpc$^{-1}$ and $\Omega_{\rm{m}} = 0.3$ this corresponds to a distance of 73.2 Mpc when accounting for peculiar velocities according to the 
NED\footnote{\href{https://ned.ipac.caltech.edu}{https://ned.ipac.caltech.edu}} infall model.

Our first ZTF photometry for SN\,2019tsf (a.k.a. ZTF19ackjszs)
was obtained on 2019 Oct 29 ($\mathrm{JD}=2458786.03$) with the P48.
The first detection is in $r$ band, with a host-subtracted magnitude of $17.40\pm0.06$~mag, at $\alpha=11^{h}08^{m}32.80^{s}$, $\delta=-10\degr28\arcmin54.4\arcsec$ (J2000.0).
This transient was first reported to the TNS by Gaia on Oct 30 \citep{2019TNSTR2225....1H}, and later also by  ATLAS, ZTF and Pan-STARRS.

The host galaxy of SN 2019tsf is NGC 3541, which has a well established redshift of
$z = 0.021$ 
and a redshift independent distance of 83.9 Mpc from \cite{2014MNRAS.445.2677S}, which we will adopt here.

\begin{figure*}
\centering
     \includegraphics[width=0.8\textwidth]{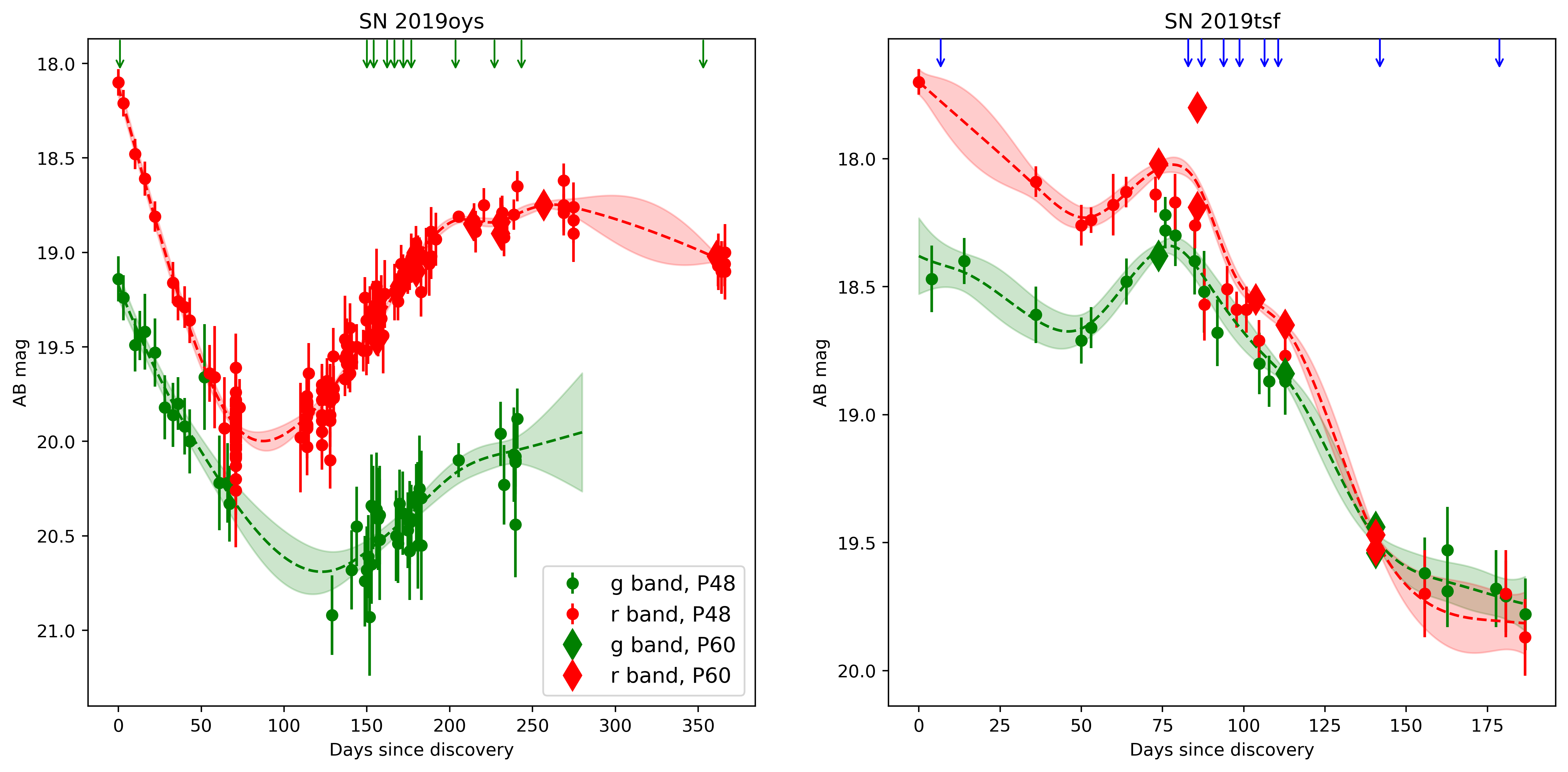}
    \caption{Light curves of SN 2019oys (left) and SN 2019tsf (right) in $g$ (green symbols) and $r$ (red) band. These are observed (AB) magnitudes plotted versus observer frame time in days since first detection. Both these Type Ib SNe showed a dramatic increase in brightness after months of decline, and in the case of SN 2019oys that rebrightening continued over more than 100 days. The arrows on top indicate the epochs of spectroscopy, and the lines with error regions are Gaussian Process estimates of the interpolated LC, which were used to absolute calibrate the spectra.
    }
    \label{fig:lcboth}
\end{figure*}

None of these transients had constraining pre-explosion detections. For SN 2019tsf, Gaia reported upper limits from August, 3 months prior to discovery, and SN 2019oys had similarly non-constraining limits from end of May. It seems that both SNe exploded when in Solar conjunction and were only discovered while already on the decline. The typical rise time for a Type Ib supernova is $\sim22$ days \citep{taddia2015}, so it is likely that we missed both the rise and the peak. Given the absolute $r$-band magnitude at discovery, they were likely found within a month from peak \citep[compare][their fig.~7]{Taddia2018}. This is also consistent with the classification spectra.
Since we do not know the time of explosion, 
throughout this paper we will always discuss both transients with phases with respect to first detection, as given above.

We classified SN 2019oys based on a spectrum obtained on 2019 Aug 29
with the Palomar 60-inch telescope (P60; \citealp{2006PASP..118.1396C}) equipped with the 
Spectral Energy Distribution Machine (SEDM; \citealp{2018PASP..130c5003B}). It was the only spectrum we obtained of this transient in 2019, and we reported the classification to TNS as a Type Ib supernova. For SN 2019tsf, the classification was done by ePESSTO+ \citep{2019TNSCR2284....1M}. They reported a Type Ib supernova close to peak brightness at a redshift of about 0.03, with no note of the NGC galaxy host.

Since both of these supernovae were found declining, no additional attention was given to them for the next $\sim100$ days, but they were photometrically monitored as part of ZTF routine observations. The interest emerged again once the light curves all of a sudden started brightening at later phases.

\subsection{Optical photometry}
\label{sec:optical}

Following the discoveries, we thus obtained regular follow-up photometry during the declining phase in $g$ and $r$ band with the ZTF camera 
\citep{dekany2020} 
on the P48.
Later on, after rebrightening started, we also obtained triggered photometry in
$gri$ with the SEDM on the P60.
Light curves from the P48 come from the ZTF pipeline \citep{2019PASP..131a8003M}.
Photometry from the P60 were produced with the image-subtraction pipeline described in \cite{Fremling2016}, with template images from the Sloan Digital Sky Survey (SDSS; \citealp{2014ApJS..211...17A}). This pipeline produces PSF magnitudes, calibrated against SDSS stars in the field. 
All magnitudes are reported in the AB system.

In our analysis we have corrected all photometry for Galactic extinction, using the Milky Way (MW) color excess 
$E(B-V)_{\mathrm{MW}}=0.06$~mag toward the position of SN 2019tsf
and
$E(B-V)_{\mathrm{MW}}=0.08$~mag toward the position of SN 2019oys
\citep{2011ApJ...737..103S}.
All reddening corrections are applied using the \cite{1989ApJ...345..245C} extinction law with $R_V=3.1$. No further host galaxy extinction has been applied, since there is no sign of any \ion{Na}{id} absorption in any of our spectra. 
The light curves are shown in Fig.~\ref{fig:lcboth}.

For SN 2019oys, the initial decline lasted at least 70 days (this is past discovery in the observer's frame). It declined quickly in the $r$ band at a rate of 3.0 mag per 100 days, and somewhat slower at 1.8 mag per 100 days in the $g$ band, thus becoming less red with time. We then have a gap in our observations, and when imaging was resumed again after about a month in December 2019, it was clear that the decline had not continued, but that in fact the light curve was now rebrightening. Once this was realized in mid-January 2020, a more intense follow-up was activated (Fig.~\ref{fig:lcboth}). A few recent $r$-band data points have been added to the light curve of SN 2019oys, which has remerged from solar conjunction and is still bright.

SN 2019tsf had a $r$-band decline over 60 days with a more normal (for SE SNe) rate of 1.4 mag per 100 days. The $g$-band light curve is more sparse, but is again shallower. For this supernova we can more clearly see the onset of the brightening after about 70 days, the $g$-band light curve rises most clearly by 0.46 mags over the next 26 days, whereas the $r$ band increases by slightly less than 0.14 mag. The light curve then peaks at 
m$^{\rm{peak}}_{r} = 18$ after 90 days after which it steadily declines again over the next 100+ days.

\subsection{Swift-observations}
\subsubsection{UVOT photometry\label{sec:uvot}}

For SN 2019oys, which did show clear evidence for CSM interaction (see below), we triggered a
series of ultraviolet (UV) and optical photometry observations
with the UV Optical Telescope onboard the Neil Gehrels
{\it Swift} observatory ($UVOT$; \citealp{2004ApJ...611.1005G}; \citealp{2005SSRv..120...95R}). 
Our first {\it Swift-UVOT} observation was performed on 2020 Mar 9 and provided detections in all the bands. However, upon inspection it is difficult to assess to what extent the emission is actually from the supernova itself, or if it is diffuse emission from the surroundings. 
The last $u$-band detection appears to be real and point-like
($u=20.16^{+0.30}_{-0.23}$ mag (AB) at 
$\mathrm{JD}=2458986.81$), 
but for the remaining bands we would need to await template subtracted images to get reliable photometry. Unfortunately, the SN was still brightening as it went behind the Sun.

\subsubsection{X-rays}
\label{sec:xrays}

With {\it Swift} we also used the onboard X-Ray Telescope (XRT; \citealt{Burrows2005}). We used online analysis tools \citep{Evans2009} to search for X-ray emission at the location of SN 2019oys. Combining the five epochs taken in March 2020 amounts to a total XRT exposure time of 12\,251 s (3.4 h), and provides a marginal detection with $16.7^{+3.5}_{-2.8}\times10^{-3}$~counts~s$^{-1}$ between 0.3 and 10 keV. However, again it is not possible to assess if this is emission from the transient or from the host galaxy.  We can conservatively treat this as an upper limit on the possible X-ray luminosity of the supernova itself. If we assume a power-law spectrum with a photon index of $\Gamma = 2$ and a Galactic hydrogen column density of $9.3\times10^{20}$~cm$^{-2}$ \citep{HI4PI2016a} this would correspond to an unabsorbed 0.3–10.0 keV flux of $7.5\times10^{-13}$~erg~cm$^{-2}$~s$^{-1}$. At the luminosity distance of SN 2019oys this corresponds to a luminosity of L$_{\rm{X}} < 4.7\times10^{41}$~erg~s$^{-1}$ at an epoch of $\sim200$ rest-frame days since discovery.

\subsection{Optical spectroscopy}
\label{sec:opticalspectra}

Spectroscopic follow-up was conducted with SEDM mounted
on the P60. Further spectra were obtained with the Nordic Optical Telescope (NOT) using the A. Faint Object Spectrograph (ALFOSC), with the Keck-I telescope using the Low Resolution Imaging Spectrograph (LRIS; \citealp{1994SPIE.2198..178O}), 
and with the Device Optimized for the LOw RESolution (DOLORES) on Telescopio Nazionale Galileo (TNG). 
A log of the spectral observations is provided in Table~\ref{tab:spec}, which includes 19 epochs of spectrosopy (9 for SN 2019tfs and 10 for SN 2019oys).
Two of the NOT spectra were obtained with a somewhat higher resolution than we normally use (grism 8 instead of grism 4) to probe the width of the narrower lines. 
These observations were taken for SN 2019oys on days 167 and 243.
The \texttt{LPipe} reduction
pipeline \citep{perleyspec}
was used to process the LRIS data. SEDM spectra were reduced using the pipeline described by
\citet{rigault} and the
spectra from La Palma were reduced using standard pipelines and procedures for each telescope and instrument. All spectral data and corresponding information will be made available via WISeREP\footnote{\href{https://wiserep.weizmann.ac.il}{https://wiserep.weizmann.ac.il}} \citep{Yaron:2012aa}.

\subsection{Radio observations}
\label{sec:radio}

Radio observations of the field of SN 2019oys were conducted using the Arcminute Microkelvin Imager - Large Array (AMI-LA; \citealt{zwart_2008}; \citealt{hickish_2018}). AMI-LA is an interferometer made up of eight 12.8 m antennas which
 operates with a 5 
GHz bandwidth around a central frequency of 15.5 GHz. We conducted our first two AMI-LA observations of SN 2019oys on Sept 19 and 23, 2019.
Initial data reduction, flagging, and calibration of the phase and flux, 
was carried out using a customized AMI data reduction software package. 
Phase calibration was done using interleaved observations of J0714+3534, while absolute flux calibration was achieved against 3C286. Additional flagging was performed using CASA. 

\begin{figure*}
\centering
\includegraphics[width=15cm]{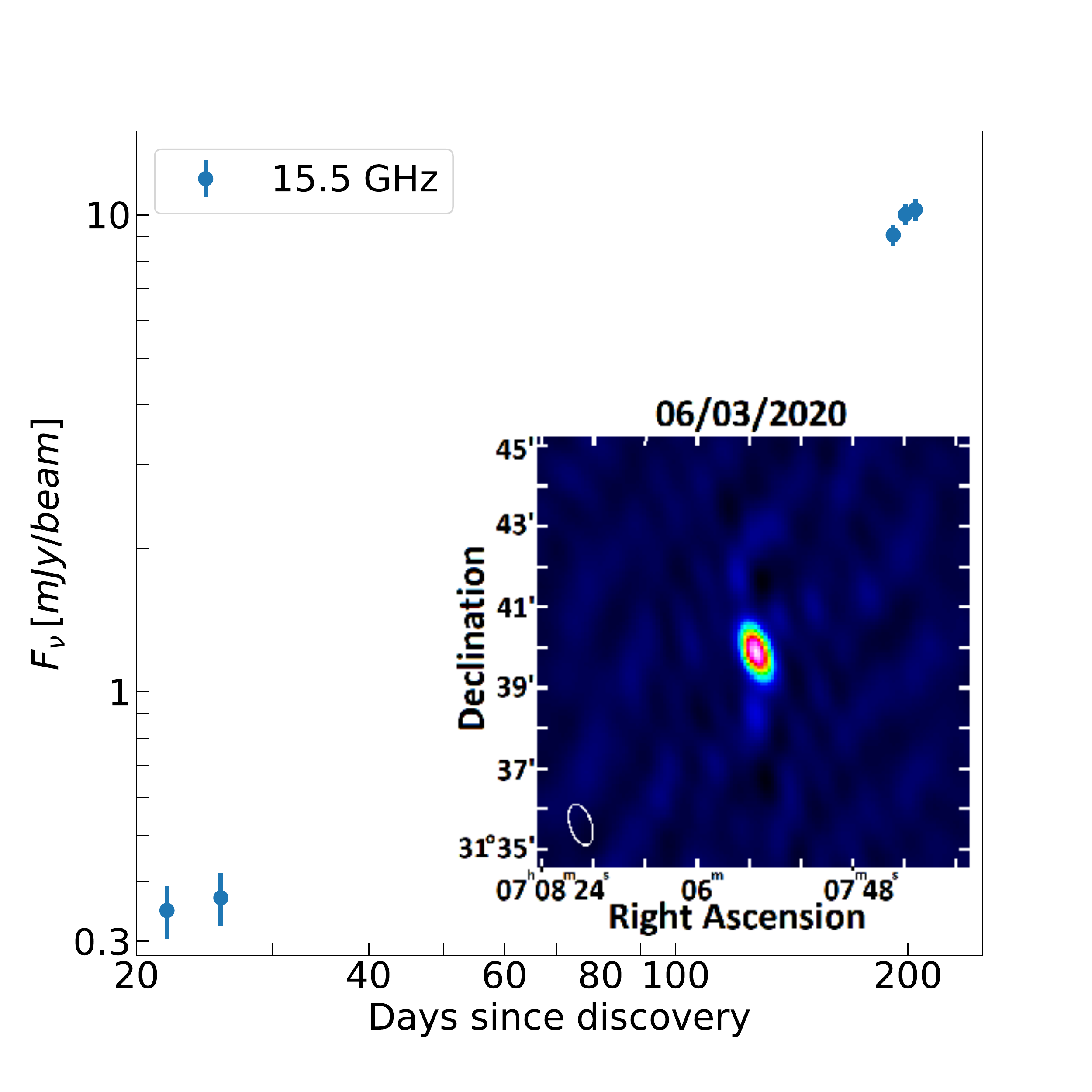}
\caption{
The radio light curve of SN 2019oys at 15.5 GHz as observed with AMI-LA. The inset in the lower right shows the radio image from Mar 6 2020, when the SN was detected at a level of 9 mJy. Fluxes are provided in Table \ref{Table: radio data}. 
}
\label{fig:radioLCanddetect}
\end{figure*}

The first two radio observations resulted in detections of a source at the phase center with an estimated flux of 0.35 mJy at 15.5 GHz, but with no apparent flux evolution. Following the spectacular coronal line spectrum obtained for SN 2019oys at the Keck telescope, providing strong evidence for CSM interaction, we triggered AMI-LA again on Mar 6 2020. This observation provided a strong detection of the SN with a significantly higher flux of 9 mJy at 15.5 GHz, and the radio image is shown in the inset of the radio light curve in
Fig.~\ref{fig:radioLCanddetect}. 

We also observed the field of SN\,2019oys with the Karl G. Jansky Very Large Array (VLA) 
on Mar 16 2020\footnote{DDT program VLA/20A-421; PI Horesh.},
while the VLA was in C configuration. The observations were performed in the S- ($3$\,GHz), C- ($5$\,GHz), X- ($10$\,GHz), Ku- ($15$\,GHz), K- ($22$\,GHz) and Ka- ($33$\,GHz) bands. 
We report here a spectral radio peak of $F_{\nu} = 21.5 \pm 1.0$\,mJy at a frequency $\nu = 23.5 \pm 1.3$\,GHz. 
The log of the radio observations and measurements is provided in 
Table~\ref{Table: radio data}.
We hope to continue monitoring SN\,2019oys with the VLA.

\section{Discussion}\label{sec:discussion}

\subsection{Light curves}

\begin{figure*}
\centering
    \includegraphics[width=0.8\textwidth]{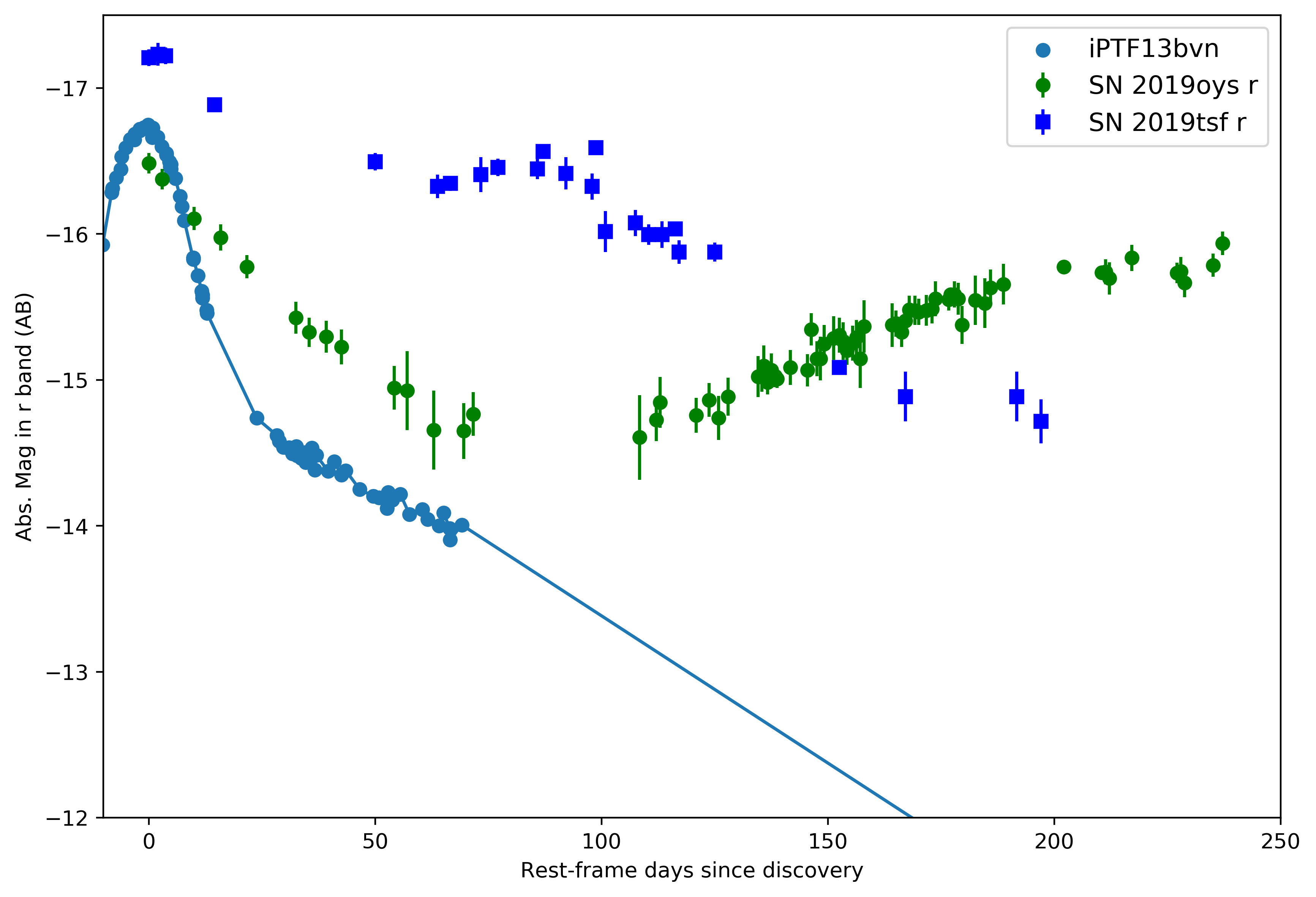}
    \caption{Light curves in absolute $r$-band magnitude 
    (M$_{\rm{r}}$) 
    for our two supernovae.  This accounts for distance modulus and MW extinction as discussed in the text, but no additional corrections for host extinction. 
    In addition we have plotted the Type Ibn SN iPTF13bvn \citep{Fremling2016}, which is a typical radioactively powered stripped envelope supernova. This SN was shifted in time to match the peak to the discovery dates of our SNe. The photometry has been binned to nightly averages.
    }
    \label{fig:lcabs}
\end{figure*}

The $g$- and $r$-band LCs of our two SNe are displayed in Fig.~\ref{fig:lcboth}.  
The general behaviour of the LCs was already discussed in Sect.~\ref{sec:optical}, and the main characteristic is of course the linear decline which is suddenly turned into a rebrightening. In Fig.~\ref{fig:lcabs} we show both LCs together in absolute magnitudes (here in the $r$ band). This shows that SN 2019tsf is more luminous than SN 2019oys by almost a magnitude at discovery, and remains brighter until about 150 days later, when the prolonged rebrightening of 
SN~2019oys catches up. 
For comparison we have also included a typical Type Ib supernova, iPTF13bv from \cite{Fremling2016}. This SN LC has been shifted by about two weeks for the maximum brightness to coincide with the discovery of our two SNe, and the distance and MW extinction have been adopted from \cite{Fremling2016}. The maximum brightness for iPTF13bvn is similar to what we see at discovery for our two SNe, but after the diffusion phase the normal Type Ib fades faster. There is more late time photometry available for iPTF13bvn, and the line in the figure connects smoothly to these data at about 200 days when iPTF13bvn is much fainter than SNe 2019oys and 2019tsf. Our SNe clearly show very different LCs, and this is further discussed in Sect.~\ref{sec:conclusions}.

We do not have enough photometric bands to construct a proper bolometric LC. We caution therefore that the strong brightening in the $r$ band for SN 2019oys is to a large extent due to line emission in H$\alpha$. The $g-r$ color got steadily bluer during the decline of the light curve, while in the rising phase the color is again quite red.
Between 150 and 172 days, H$\alpha$ increased from $\sim 60 \%$ to  $\sim 72 \%$ of the $r$-band flux. 
This is reminiscent of the LC of the Type IIn SN 2006jd, where the $r-$band flux reached a minimum at $\sim 190$ days, and then again brightened by $\sim 1$ mag.
The quasi-bolometric light curve of SN 2006jd
showed a flat behaviour and later a decline during this period \citep[][their fig. 9]{stritzinger05ip}. A difference between SN 2019oys and SN 2006jd is that the dip in the $r$ band is more shallow and the minimum occurs at a later epoch
for SN 2006jd.

\subsection{Spectroscopy}

\begin{figure*}
\centering
\includegraphics[width=12cm,angle=90]{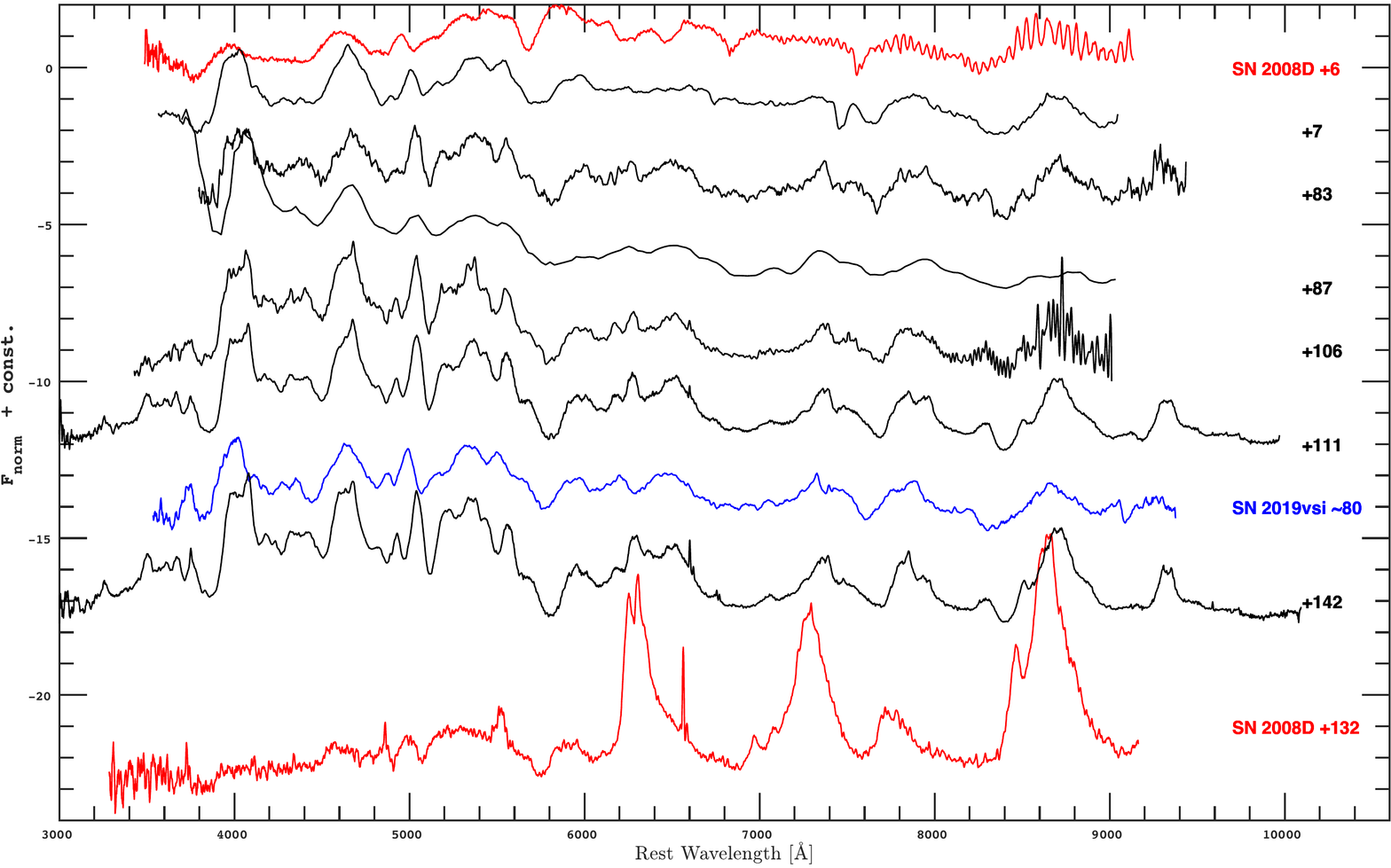}  
\caption{\label{fig:specszs} The spectral sequence of SN 2019tsf demonstrates that the spectral evolution is quite slow. We show a selection of the spectra listed in Table~\ref{tab:spec}. Phases given in rest-frame days are provided for each spectrum. The uppermost spectrum  (in red) is of the Type Ib SN 2008D obtained 6 days past maximum light from \cite{malesani2008D}. This 
gives the best match of the classification spectrum using SNID. The spectra obtained at $\sim100$ days when the supernova was rebrightening are still quite similar to the typical Type Ib SN spectrum obtained close after discovery. 
The third to last spectrum (in blue) is of the Type Ib SN 2019vsi about 80 days past discovery, 
and shows great similarity with the spectra of SN 2019tsf.
In addition, and as a an example of a more normal late time spectrum we also show SN 2008D at t = 132 d presented in \cite{Modjazetal2009}. 
SE SNe at these epochs are typically dominated by forbidden emission lines of \ion{Mg}{I}], [\ion{O}{I}], and [\ion{Ca}{II}].
No signs of narrow lines or other features signalling CSM interaction can be found in the spectra of SN 2019tsf, in stark contrast to the case of SN 2019oys.
The spectra are normalized and offset for clarity. All data will be made available via WISEREP.
}
\end{figure*}

For SN 2019tsf the classification spectrum revealed a Type Ib supernova (\citealt{2019TNSCR2284....1M}; Sect.~\ref{sec:detection}). When we run SNID \citep{Blondin2007} on this spectrum, the best match is SN~2008D, a well monitored Type Ib. We show this comparison in Fig.~\ref{fig:specszs}. The next spectrum was only obtained more than two months later, after the brightening, with the NOT using ALFOSC (Table~\ref{tab:spec}).
The aim of this second spectrum was to search for evidence for CSM interaction that could explain the rising light curve. 
As can be seen in Fig.~\ref{fig:specszs}, 
no signatures of CSM interaction are present in the spectra.
We continued the spectroscopic campaign with spectra from P60, 
TNG, Keck and NOT - until the SN faded out of spectroscopic sight. The spectral evolution was quite slow - no significant evolution is apparent in the sequence from 80 to 180 days from discovery. In Fig.~\ref{fig:specszs} we also compare the late spectra of SN 2019tsf with that of another ZTF supernova, the Type Ib SN 2019vsi. That spectrum was 
obtained with NOT+ALFOSC about 80 days past discovery and shows great similarity to the spectra of SN 2019tsf. 
Overall, we see little spectroscopic evidence that SN 2019tsf interacted with a CSM.

\begin{figure*}
\centering
\includegraphics[width=12cm,angle=90]{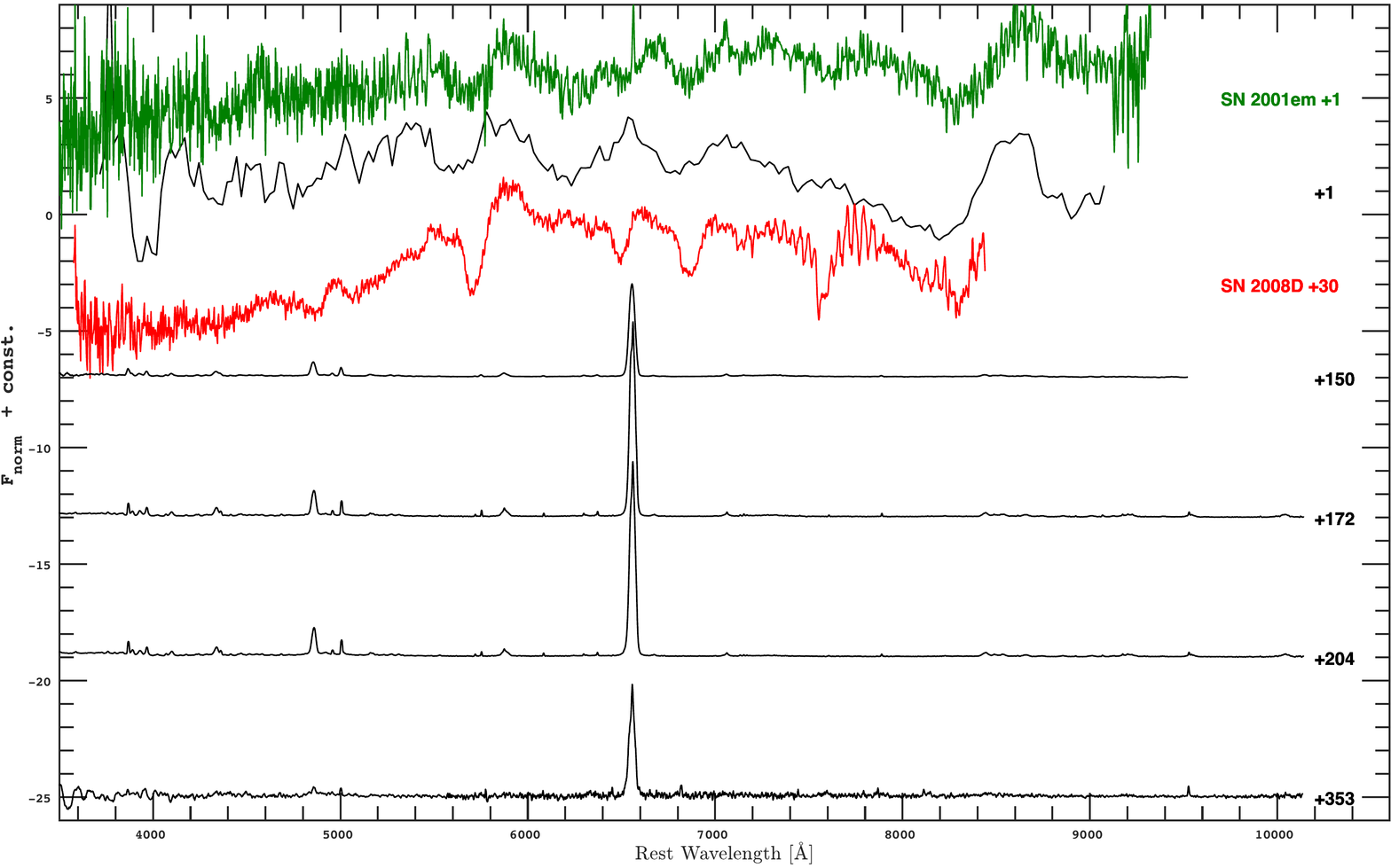}  
\caption{ \label{fig:specwzt} 
The spectral sequence of SN 2019oys shows an abrupt change from the very first Type Ib spectrum obtain by the P60, to the later spectra acquired once the light curve started to brighten. These latter spectra show clear evidence for CSM interaction as evidenced by the dominance of the narrow emission lines. We show a selection of the spectra listed in Table~\ref{tab:spec} for this supernova. Phases in rest-frame days are provided for each spectrum. The third spectrum from the top is of the Type Ib SN 2008D
\citep{malesani2008D} at 30 days past maximum light (red), which gives the best match to our classification spectrum using SNID.
In addition, we compare to the +1 days spectrum of SN 2001em from \citet{Shivvers+2019} (green).
The spectra obtained at $\gtrsim150$ days when the supernova was re-brightening are quite similar to the spectra of the Type IIn SN\,2015ip and this is highlighted in Fig.~\ref{fig:specwztlog}.
}
\end{figure*}

\begin{figure*}
\centering
\includegraphics[width=12cm,angle=90]{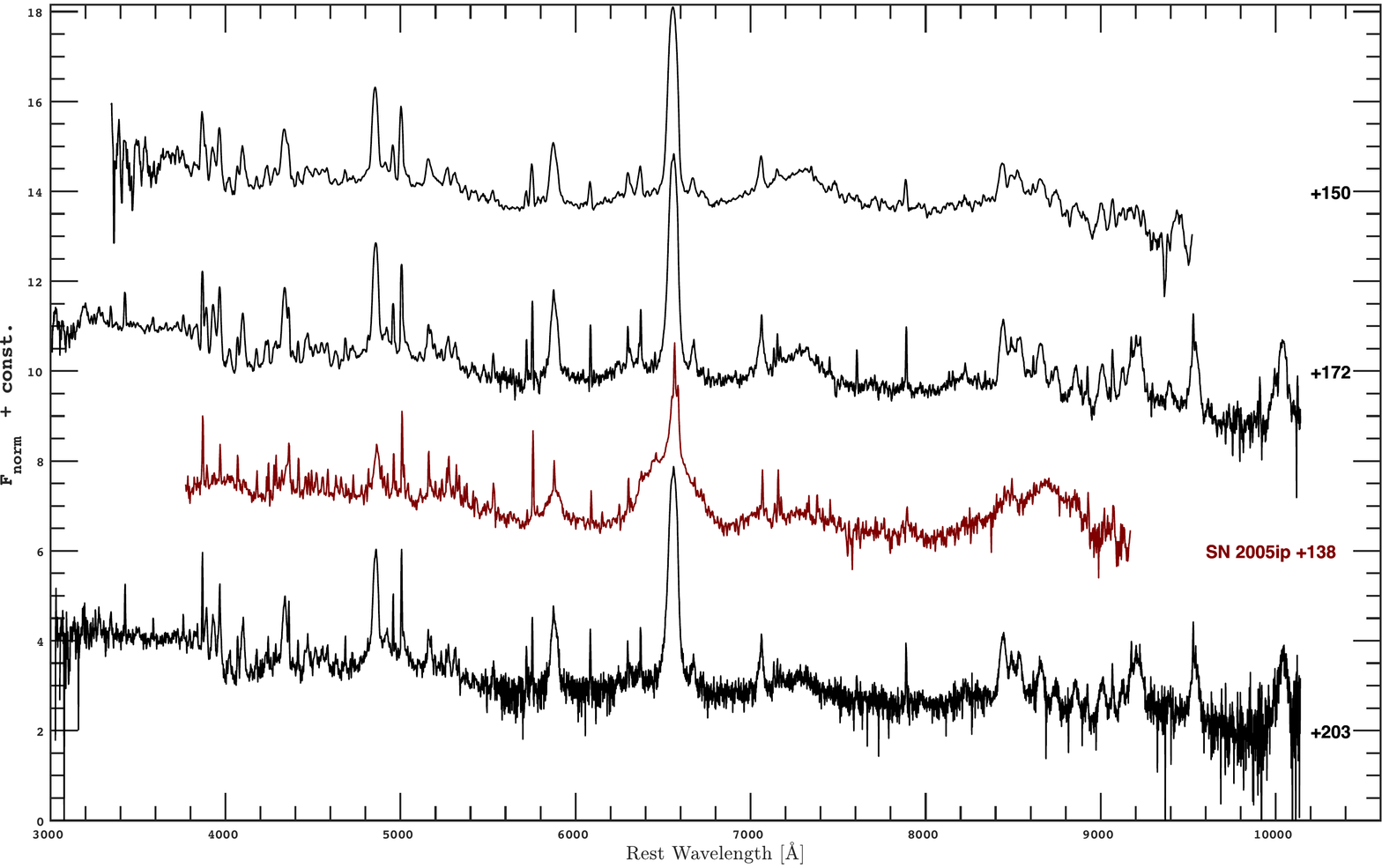}  
\caption{\label{fig:specwztlog} Spectral sequence of SN 2019oys during rebrightening. A handful of the spectra listed in Table~\ref{tab:spec} are shown, and here on a logarithmic scale to highlight the bright narrow emission lines. Phases in rest-frame days are provided for each spectrum. 
A spectrum of the Type IIn SN 2015ip is shown for comparison (in red). This spectrum is from \citet{stritzinger05ip} taken at 138 days past discovery.
Basically all high excitation coronal lines seen in SN 2005ip are also detected in SN 2019oys, a main difference being that our SN do not display the broad H$\alpha$ line from hydrogen-rich ejecta.
The spectra are normalized and offset for clarity. 
}
\end{figure*}

On the contrary, SN 2019oys displayed a spectacular metamorphosis. The first classification spectrum displayed a Type Ib SN with no signs of CSM interaction. Again, SN 2008D provides the best match by SNID, as illustrated in Fig.~\ref{fig:specwzt}. That spectrum of SN 2008D was obtained 6 days past peak brightness, and is again an indication that our SNe were discovered past peak, but not by much. 
Also SN 2019oys was basically ignored for a long time, it was not considered interesting enough for spectroscopic follow-up given the lack of a 
well-determined explosion date.  When we realised the supernova was on the rise, we triggered 
the NOT, which revealed a booming narrow-line dominated spectrum. 
This was completely unlike the first spectrum. Wondering whether we might have missed some of these narrow features in the early very low dispersion SEDM spectrum, we took another SEDM spectrum just a few days later - and again got an emission line dominated spectrum, with a particularly strong H$\alpha$ line. Whereas the SEDM spectra cannot reveal the dense forest of narrow lines, the metamorphosis was clearly apparent also in this comparison.  The spectral sequence displayed in Fig.~\ref{fig:specwzt} 
illustrates 
this; the sudden transition from a Type Ib to what is better described as a Type IIn supernova.
To properly showcase the evolution of the spectra on the re-brightening part of the light curve, we show these spectra on a logarithmic scale in Fig.~\ref{fig:specwztlog}. This allows displaying our best observations showing a sequence of  dense narrow-line spectra rich in high-ionization coronal lines. 
This figure also includes a comparison with a spectrum of the spectacular coronal line supernova SN 2005ip, taken from \cite{stritzinger05ip}. 
SN 2005ip was a 
supernova that displayed many similarities to SN 2019oys. It was first classified as a Type II supernova, although in hindsight it did display a number of narrow emission lines already close to discovery. The light curve of SN 2005ip also lacked constraints on the time of explosion, but did after about 200 days stop declining and entered more of a flat plateau, rather than the more dramatic increase in brightness that SN 2019oys delivered. The richness of coronal lines in SN 2005ip was unprecedented
\citep{smith2009}. 
The spectrum of SN 2019oys is equally rich, and we provide a list of line identifications in 
Table~\ref{tab:coronallines}. The supernova 
displays high ionization species such as for example
[\ion{Ar}{xiv}],
[\ion{S}{xii}],
[\ion{Ne}{V}] and
[\ion{Fe}{XI}].
We provide our line identifications on the merged day 172 and 204 spectrum of SN 2019oys in 
Fig.~\ref{fig:lineid}. 

We measured the lines with Gaussian fits using {\tt iraf/splot} on the Keck spectrum from day 172. This was combined with the spectrum from day 204 to increase the signal to noise since these two spectra were virtually identical, but the flux scale was set to that of the day 172 spectrum. We used the nearby continuum as the baseline, and slightly varying the position of the continuum provides an estimate on the uncertainty in the emission line flux. For example, the [\ion{Ne}{V}] $\lambda$3426 has a flux of $5.4\times10^{-16}$~erg~s$^{-1}$~cm$^{-2}$ with an uncertainty of less than $7\%$. However, we caution that whereas this provides a good measure of the significance of the line, using such fluxes for diagnostics is better done in connection with a physical model that can justify the real baseline, and this is well illustrated in Fig.~\ref{fig:synthetic}. The line measurements are provided in Table~\ref{tab:coronallines}. Note that this is only a selection of the numerous lines, the spectra are made available for further measurements. The narrow lines are unresolved at the spectral resolution (7 \AA ) of LRIS with grisms 400/3400/8500.

Table~\ref{tab:coronallines} demonstrates that we detect and identify most of the multitude of emission lines also detected in SN 2005ip. Some notable exceptions are the 
[\ion{S}{ii}]~$\lambda\lambda 6717,6731$ that were strong in SN 2005ip, but are very weak in SN 2019oys. 
Overall, however, the conditions present in the line forming region(s) must be quite similar between these two supernovae, these conditions were also studied in detail by \cite{stritzinger05ip} and earlier also for the Type IIn SNe 1995N \citep{fransson95N}
and 2010jl \citep{fransson10jl}. We discuss some diagnostics of the emission lines in the subsections below.

There are of course also some differences between the two above-mentioned SNe. Figure~\ref{fig:specwztlog} shows that SN 2005ip displayed a broad component of H$\alpha$, which is not present in SN 2019oys. This is a signature of the hydrogen-rich fast-moving ejecta that this Type II SN showed already from early times. SN 2019oys is instead a stripped envelope Type Ib SN, and such a supernova is less likely to metamorphose into a rich coronal line dominated transient.

In fact, less than a handful of SE SNe are known to have transitioned to CSM interacting objects, as mentioned in the introduction.
SN 2017ens \citep{2018ApJ...867L..31C} was a very/super luminous Type Ic-BL that hit CSM after 150 days. It also displayed some coronal lines, but the light curve 
never rebrightened. This was a unique object, but indeed shares many properties with SN 2019oys. 
SN 2017dio \citep{2018ApJ...854L..14K} was a Type Ic that already from the start showed evidence for CSM interaction in terms of narrow emission lines. Rather than showing a spectacular change in spectral properties, it displayed a double nature with pseudo-continuum Type Ic spectral features with narrow Balmer lines 
$\grave{\rm{a}}$ la
Type IIn superimposed. 

Finally, we must mention SN 2014C \cite[e.g.,][]{Milisavljecic2015} which is a well studied SE SN that ran into CSM and 
which transformed from a Type Ib to a Type IIn SN, just like SN 2019oys, a phenomenon also seen in 
SN 2001em \citep[][]{ChugaiChevalier, poonam2020}.
Also in this case, late-time high-resolution spectra revealed coronal lines. This small family of changing type SNe demonstrates the existence of nearby dense hydrogen-rich CSM close to - but not too close to - the stripped progenitor star, possibly from binary evolution and/or violent eruptions 
\citep[e.g.,][]{2020arXiv200309325S}. Recently, SN 2018ijp was also interpreted as a SE SN with \say{delayed interaction} \citep{2020arXiv200903331T}.

Having mentioned several SNe showing a similar spectroscopic transition as did SN 2019oys, it is worth reminding the reader that the LC of SN 2019oys is quite unique within this sample. Whereas SN 2005ip displayed a drastic change in decline when the CSM interaction started in earnest, the late light curve was more of a plateau than an actual rise. SNe 2014C, 2017dio and 2017ens also did not show signs of rebrightening, although for SN 2017dio it is possible that we missed the early phases.

\subsubsection{Lyman-alpha fluorescence}
\label{sec:lya}

The near-infrared (NIR) part of the spectrum shown in Fig.~\ref{fig:lineid} displays a strong complex of lines.
This has been seen previously in other SNe, in particular in SN 1995N \citep{fransson95N} and also for SN 2005ip 
\citep{Fox2020}
and has been explained as the result of fluorescence of \ion{Fe}{II} by Ly$\alpha$. In this process, electrons in the $a^4D^e$ 
level 
of \ion{Fe}{II} are excited to levels $\sim 11.2$\,eV above the ground state by accidental resonances with the Ly$\alpha$ line \citep{johansson84,sigut98,sigut03}. The cascade to lower levels results in a UV line and a line in the NIR.  Because these levels with high excitation temperatures are difficult to excite by thermal collisions, the presence of these NIR 
lines is a strong signature of radiative pumping by Ly$\alpha$. Although we do not have any UV spectra, especially our Keck spectra allow us to  examine the NIR features. 

Because of the multitude of Fe II lines all over the optical and NIR ranges, spectral simulations are required in order to identify the most likely and strongest transitions. 
We use the predicted line fluxes from the AGN simulations by  \citet{sigut03}. 
The relative fluxes of all Fe II lines from the list of \citet{sigut03} are shown as vertical bars in 
Fig. \ref{fig:synthetic}. 
While the model identifies most of the  Fe II lines in the range below
$\sim5300$~\AA,
which can be thermally excited, the interesting region is 
at $\sim8400-9600$~\AA, as can be seen in the lower panel of Fig.~\ref{fig:lineid}. 
In addition to the broad Paschen lines up to at least the $n=14\rightarrow3$ transition, the \ion{Ca}{II} triplet, and narrow [\ion{S}{III}] $\lambda\lambda 9069,9531$ lines, there are also a number of narrow lines from \ion{Fe}{II}. While several of these are blended with lines from other ions, there are some lines which are not coming from lighter elements. In particular, the lines at   
8927,  9123, 9132, 9176, 9178~\AA~can 
not be identified with other ions. These, together with the \ion{Fe}{ii} $\lambda 8451$ line, are also the ones expected to be strong in the model. The latter line is blended with a broad feature. While there is some contribution from high order Paschen lines, there is likely to be a strong contribution from \ion{O}{I} $\lambda 8446$, which is expected as a result of fluorescence with Lyman $\beta$ \citep{Bowen1947}. 
There is thus strong evidence for narrow \ion{Fe}{II} lines excited by Ly $\alpha$.

\subsubsection{Synthetic spectrum}
\label{sec:model}
To infer some basic properties of the CSM  we have also calculated a synthetic spectrum  of SN 2019oys. This was also very useful to help in the identification of the lines in view of the line blending and many \ion{Fe}{II} lines present. In this analysis we follow the method in 
\citet{Fox2020},
and here only summarize the main ingredients.

The analysis assumes a two-zone model with separate densities for the narrow-line region and for the region responsible for the broader, electron scattering affected lines. Temperature and densities of these zones are treated as parameters. Model atoms including both collisional and radiative rates are calculated for \ion{H}{I},  \ion{He}{I},  \ion{N}{II},  \ion{O}{I},  \ion{O}{III},  \ion{Ne}{III-V},  \ion{Ca}{II} and  \ion{Fe}{VI-VII}.  As shown in Sect.~\ref{sec:lya}, \ion{Fe}{II} is strongly affected by line fluorescence by Ly$\alpha$, and would require a much more sophisticated treatment. Instead we rely on the calculation by \cite{sigut03}, which although intended for AGN conditions, should give approximate fluxes for the strongest lines, helping in identification of the multitude of \ion{Fe}{II} lines.

\begin{figure*}
\centering
\includegraphics[width=1.0\textwidth]{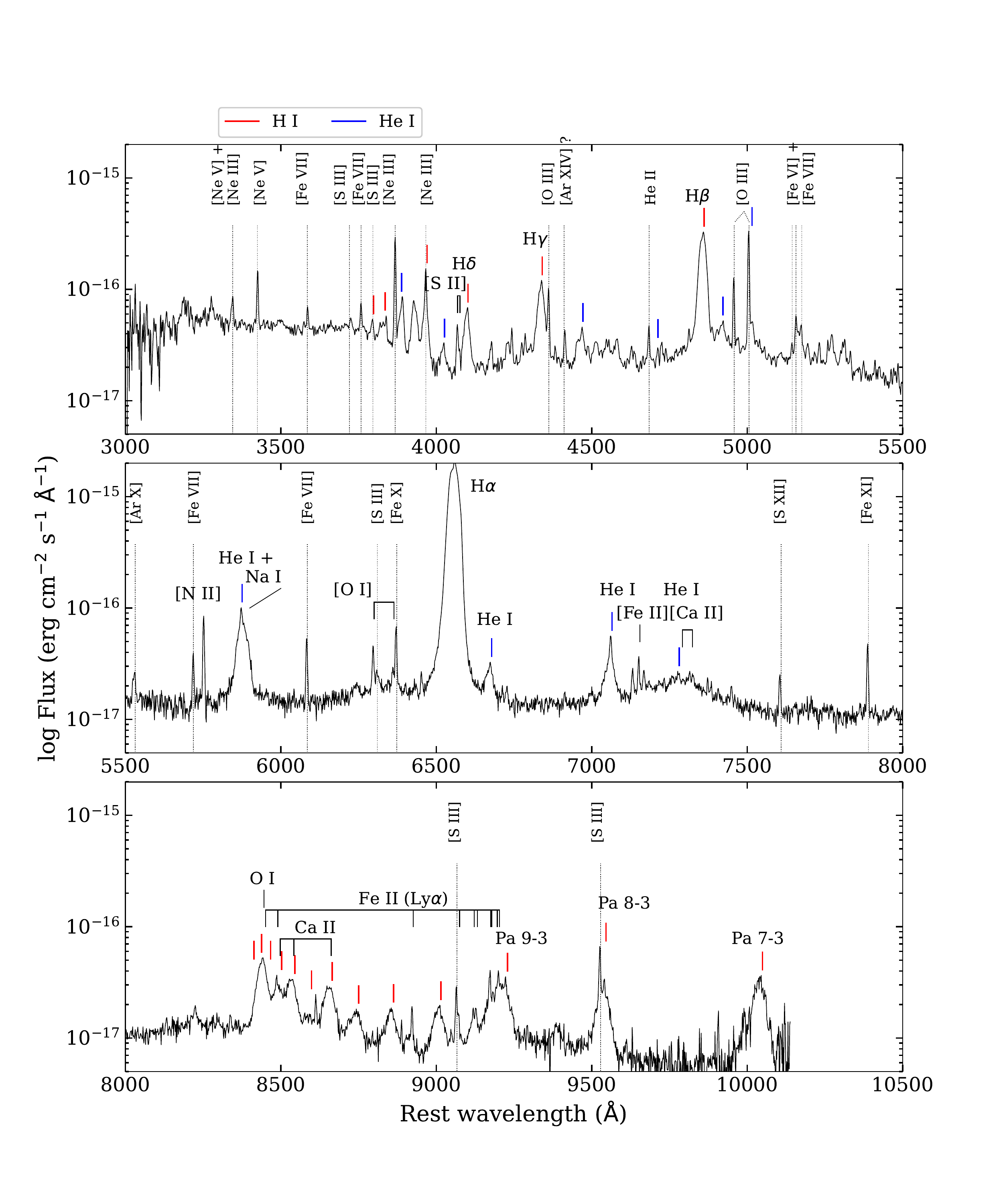}
    \caption{Line identifications of the reddening corrected, merged spectrum  of SN 2019oys from days 172 and 204. The high ionization narrow lines are marked at the top. The \ion{H}{I} and  \ion{He}{I} lines are marked with red and blue lines, respectively.
    }
    \label{fig:lineid}
\end{figure*}

\begin{figure*}
\centering
\includegraphics[width=1.\textwidth]{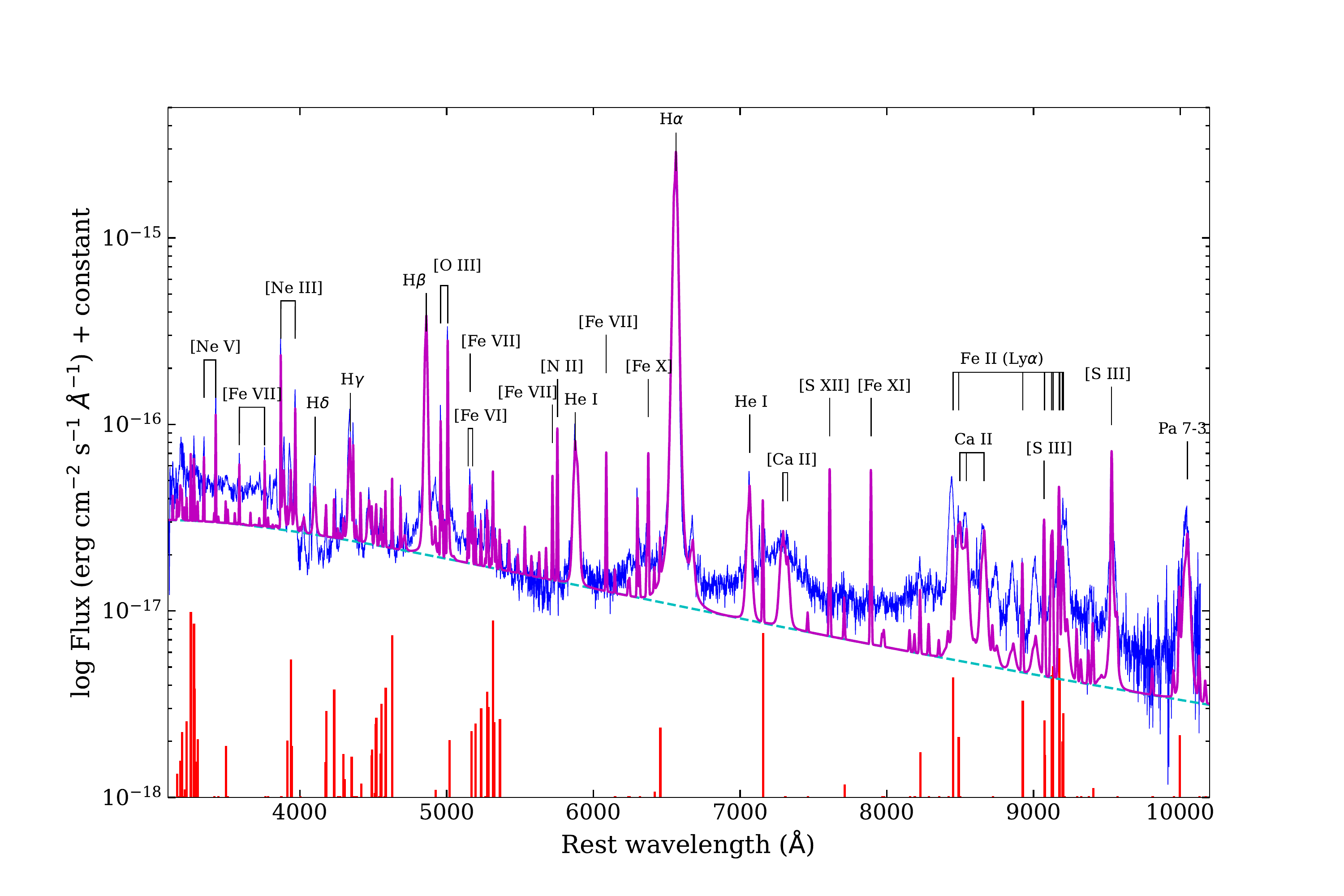}
    \caption{Synthetic spectrum (magenta) together with the reddening corrected, merged spectrum  of SN 2019oys from days 172 and 204 (blue) with line identifications. The assumed blackbody continuum is shown as the dashed cyan line. The red marks at the bottom indicates the forest of \ion{Fe}{II} lines with heights reflecting the relative intensities adopted from \cite{sigut03}. These provide indications for where to expect contribution from iron lines, and which regions are not expected to be contaminated by such emission.}
    \label{fig:synthetic}
\end{figure*}

The temperature of the CSM was set to 20,000~K, and we have assumed a blackbody background continuum with a temperature of 9500~K. As shown in \cite{Dessart:aa}, the real spectrum is likely to depart from a blackbody, especially in the UV, but lacking more accurate calculations we use a blackbody as 
an approximation.  For the line profiles we assume a Gaussian shape for the narrow high-ionization lines and an exponential line profile for the broad lines affected by electron scattering \citep{2018MNRAS.475.1261H}. For H$\alpha$ this is seen to extend to at least $\sim \pm4000$ km~s$^{-1}$. 
The synthetic spectrum is displayed in Fig.~\ref{fig:synthetic}, together with the reddening corrected, merged  spectrum  of SN 2019oys from days 172 and 204 (Fig. \ref{fig:lineid}).

Starting with the broad electron-scattering dominated lines, the steep Balmer decrement with F(H$\alpha$)/F(H$\beta) \sim 8.5$ requires a high density to produce optically thick Balmer lines. The density depends on the temperature, and assuming 20,000 K requires a density of the broad line region of $\sim 2 \times 10^9 \ \rm cm^{-3}$. While this gives a good agreement for the H$\alpha$/H$\beta$ ratio  and the unblended Paschen lines, it under-produces the higher Balmer lines by a factor of $\sim 2$. This is most likely a limitation of the one-zone model for the broad lines. In reality, the higher Balmer lines may arise deeper as compared to H$\alpha$, where the density and temperature is higher. The relative fluxes of the \ion{He}{I} $\lambda \lambda 5876,  6678, 7065$ lines are also well reproduced. The  \ion{He}{I} $\lambda 5876$ line has a different shape compared to the other  \ion{He}{I} and  \ion{H}{I} lines. This is well explained as a result of blending with the  \ion{Na}{I}~D doublet, thus confirming the presence of these lines. In the NIR, the  \ion{Ca}{II} triplet is part of the feature at $8500 - 8650$ \AA, consistent with broad lines.

For the narrow line region we assume a density 
$10^5 - 10^6 \ \rm cm^{-3}$. 
The lower density agrees with the large  [\ion{O}{I}]  $\lambda \lambda 6300, 5577$ ratio. The [\ion{O}{III}] $\lambda \lambda 5007, 4363$, [\ion{S}{III}] $\lambda \lambda 9531, 6312$ and [\ion{Fe}{VI-VII}] lines, on the other hand, indicate a higher density, $\sim 10^6  \ \rm cm^{-3}$ is needed in order to explain the auroral lines. A similar density is also indicated by the  [\ion{Ne}{III}] $\lambda \lambda 3869, 3342$ lines, although the latter line is blended with [\ion{Ne}{V}] $\lambda 3346$. From the known ratio of [\ion{Ne}{V}] $\lambda 3426$ to [\ion{Ne}{V}] $\lambda 3346$ we can, however, determine an approximate flux of the  [\ion{Ne}{III}] $\lambda 3342$ line. 
Both the \ion{H}{I} and \ion{He}{I} have also narrow components, coming from the same region as the high ionization lines.

The main disagreement in the fit are the broad features at $\sim 7300$ \AA\ and  $\sim 5000$ \AA. The former is centered around the [\ion{Ca}{II}] $\lambda \lambda 7291, 7324$ lines. If this emission originates at deeper optical depth than the Balmer lines, the line profile could undergo more scatterings, resulting in broader wings without a central, narrow component. This may also explain the broad feature around  $\sim8600$ \AA, centered at the \ion{Ca}{II} triplet.

Summarizing, we 
find 
evidence for two different density components. One inner component with high density and large column densities, 
responsible for 
the broad, low ionization lines which are affected by electron scattering. The lower density component is instead the site where the narrow high ionization lines are formed. The latter is unaffected by electron scattering, arguing for it being exterior to the high density region. An alternative may be an anisotropic geometry, where the broad component may arise from the interaction with the ejecta and  e.g. a torus-like, dense CSM. A problem for 
such a scenario is that we do not see any evidence for macroscopic high velocity ejecta, although the  orientation may hide this.

The good match of the synthetic spectrum has guided us in the line identifications provided in Figures~\ref{fig:lineid},~\ref{fig:synthetic} and in Table~\ref{tab:coronallines}. This is particularly important in view of the many blends and 
\say{contamination}
by especially \ion{Fe}{II} lines.

\section{Interpretation and Conclusions}\label{sec:conclusions}

In this paper we have presented two SE Type Ib SNe whose light curves after months of decline suddenly started rebrightening.
Such events are no doubt rare, and these discoveries heavily rely on the sky survey of ZTF which can not only discover many different kinds of transients, but also monitor them
routinely enabling us to unravel unusual behaviour also at later epochs.
One of the supernovae, SN 2019tsf, brightens only for a month, and then returns to a declining phase of the light curve. Even though we managed to obtain high quality optical spectra at the time of the light curve bump, no clear spectral signatures of CSM interaction were seen.
SN 2019oys, on the other hand, continued to rise, 
and the spectral metamorphosis is second to none 
of the few similar changing-type SE SNe known. The CSM interaction is evident and obvious and provide us with a plethora of diagnostic lines to investigate the surrounding environments.
The dichotomy illustrated by this pair of SNe highlights a number of issues in contemporary supernova studies.

There is in fact an under-abundance of studies of what spectral signatures CSM interaction should provide.
In the context of SLSNe-I, extraordinary luminous hydrogen-free transients, CSM interaction has been considered unlikely. This was partly 
based on the lack of narrow emission lines in their spectra \citep[][]{mazzali16}, although that particular work focused more on 
modeling spectra without interaction rather than demonstrating that interaction could happen without spectral signatures. 
There are also SE SNe where interaction became more evident at late stages, 
but where narrow emission lines never dominated the spectrum \cite[e.g.,][for SN 1993J]{Matheson2000}. 
The discussion was exacerbated with the curious iPTF14hls, a Type II supernova with a spectacularly long-lived light curve \citep{2017Natur.551..210A} where 
most of the run of the mill explanations for powering mechanisms did not work out, and where CSM interaction was probably
the last scenario standing \citep{Sollerman2019}. \cite{2018MNRAS.477...74A} explained the fact that such a CSM interaction did not reveal itself
in the spectral evolution as due to a particular geometry hiding the interaction site, although actual modeling of such a mechanism remain unexplored. It is somewhat inherent in the problem that while we can do detailed diagnostics of the properties of the gas emitting the narrow lines, as illustrated in Sect.~\ref{sec:model} for SN 2019oys, it is more difficult to deduce the reasons for not seeing such emission for objects lacking the diagnostic lines such as SN 2019tsf.

Possible explanations for the lack of circumstellar lines includes that the ionization is so high that all ions emitting in the optical band are ionized. 
A high ionization may 
be the result of either a high X-ray luminosity or a low density. 
Another possibility is that in order to see narrow lines 
there has to be a large enough region of low  density material. A region with a large optical depth to electron scattering might wash out the lines completely into the continuum.
We echo the conclusions of \citet{chatzo2012,chatzo2013} in that
more detailed modeling is needed to investigate such scenarios.

Whereas the powering scenarios required to sustain long-lived or superluminous light curves without displaying conspicuous spectral signals have been discussed in the literature, the problem is somewhat intensified by the two SNe presented in this work. They (re-)brighten significantly at late times, and it is quite challenging to envision any mechanism other than CSM interaction responsible for this behaviour. The well-monitored re-brightening allowed spectroscopic observations at the time of the interaction. We are 
left with two stunningly different spectral signals - one CSM interaction scenario showing a loud and clear Type IIn spectrum while the other simply does not. This  reinforces the need for better understanding of the CSM scenario.

Returning to the light curves, and comparing to a prototypical SN Ib, such as iPTF13bvn in Fig.~\ref{fig:lcabs}. On the one hand, the peak absolute magnitude of iPTF13bvn is in the same ball-park as the brightest points for our two SNe. Note again that this is not a bolometric LC, but in the $r$ band. For iPTF13bvn, there were enough data to build and model a bolometric LC, and the conclusion was that it could be powered by 0.072 \msun~of $^{56}$Ni \citep{Fremling2016}.
If we were to power the LCs of our SNe in the same way we would need more radioactive material. Assuming for example that we just missed the diffusion peaks of the SNe, we can match their LCs to that of iPTF13bvn by shifting them.
Matching to the LC at about 50 days would require 0.2 and 0.6 \msun of $^{56}$Ni, respectively, for the two supernovae, but 
the SN LCs could also have been affected by CSM powering already on the initial fading part.

In some sense, the interpretation of SN 2019oys in terms of CSM interaction as provided here puts it in the family of well explored SNe such as SNe 2015ip and 1988Z, and the formation of the coronal lines and the luminosity of the light curve can be understood in that context. However, this leaves open several fundamental questions, since SN 2019oys was not the explosion of a hydrogen-rich progenitor forming a Type II SN. Instead it was initially classified as a Type Ib, which is more similar to e.g., SN 2014C. \citet{Milisavljecic2015} discussed three different scenarios for the origin of such a CSM; a brief Wolf-Rayet phase, eruptive ejection or confinement of CSM by surrounding stars. 
There is now ample evidence that CC SN progenitors often experience large eruptive ejections close to the time of core collapse, a phenomenon that has been anticipated, but for rather special cases \citep{woosley2007PI}.
There are many similarities between SN 2014C and SN 2019oys - like the coronal line spectrum and the FWHM of the intermediate width lines ($\sim1500$~km~s$^{-1}$). 
SN 2014C also displayed strong radio emission picked up by AMI-LA \citep{Anderson2017}, even if the emission from SN 2019oys was 10 times brighter and still rising at 200 days when we obtained our latest radio observation.
We also note the difference in that SN 2014C showed a nebular spectrum with sawtooth shaped broad emission lines from the underlying ejecta, which is less obvious in the narrow line dominated spectrum of SN 2019oys. Investigating more of these systems will help us
understand why some stripped stars engage in CSM interaction (while most do not) and why some reveal this conspicuously as did SN 2019oys, whereas others, like SN 2019tsf only provide a LC bump with no spectral CSM interaction signatures.

\begin{acknowledgements}
We thank Brad Cenko for guidance on the SWIFT data. 
David Titterington helped with AMI data and made corrections to the manuscript.
Based on observations obtained with the Samuel Oschin Telescope 48-inch and the 60-inch Telescope at the Palomar Observatory as part of the Zwicky Transient Facility project. 
ZTF is supported by the National Science Foundation under Grant No. AST-1440341 and a collaboration including Caltech, IPAC, the Weizmann Institute for Science, the Oskar Klein Center at Stockholm University, the University of Maryland, the University of Washington, Deutsches Elektronen- Synchrotron and Humboldt University, Los Alamos National Laboratories, the TANGO Consortium of Taiwan, the University of Wisconsin at Milwaukee, and Lawrence Berkeley National Laboratories. Operations are conducted by COO, IPAC, and UW. 
This work was supported by the GROWTH project 
\citep{2019PASP..131c8003K}
funded by the National Science Foundation under PIRE Grant No 1545949. 
The Oskar Klein Centre was funded by the Swedish Research Council.
Gravitational Radiation and Electromagnetic Astrophysical Transients (GREAT) is funded by the Swedish
Research council (VR) under Dnr 2016-06012. 
CF is supported by the Swedish Research Council and Swedish National Space Board.
Partially based on observations made with the Nordic Optical Telescope, operated by the Nordic Optical Telescope Scientific Association at the Observatorio del Roque de los Muchachos, La Palma, Spain, of the Instituto de Astrofisica de Canarias. Some of the data presented here were obtained with ALFOSC. 
Some of the data presented herein were obtained at the W. M. Keck
Observatory, which is operated as a scientific partnership among the
California Institute of Technology, the University of California, and
NASA; the observatory was made possible by the generous financial
support of the W. M. Keck Foundation. 
This paper is partly based on observations made with the Italian Telescopio Nazionale Galileo (TNG) operated on the island of La Palma by the Fundaci{\'o}n Galileo Galilei of the INAF (Istituto Nazionale di Astrofisica) at the Spanish Observatorio del Roque de los Muchachos of the Instituto de Astrofisica de Canarias. 
The SED Machine is based upon work supported by the National Science Foundation under Grant No. 1106171. 
We thank the staff of the Mullard Radio Astronomy Observatory, 
  University of Cambridge, for their support in the maintenance, and
  operation of AMI. We acknowledge support from the European Research
  Council under grant ERC-2012-StG-307215 LODESTONE. A.H. is grateful for the support by grants from the Israel Science Foundation, the US-Israel Binational Science Foundation, and the I-CORE Program of the Planning and Budgeting Committee and the Israel Science Foundation. We thank the referee for suggestions for how to improve Figures 4 and 5.
\end{acknowledgements}

\clearpage

\bibliography{arXiv}

\clearpage

\begin{deluxetable}{lccc}
\tablewidth{0pt}
\tabletypesize{\scriptsize}
\tablecaption{Summary of Spectroscopic Observations \label{tab:spec}}
\tablehead{
\colhead{Object} &
\colhead{Observation Date} & 
\colhead{Phase} & 
\colhead{Telescope+Instrument} \\
\colhead{} &
\colhead{(YYYY MM DD)}  & 
\colhead{(Rest-frame days)} &
\colhead{} 
}
\startdata
SN\,2019tsf & 2019 Nov 05 & 6.71 & NTT+EFOSC2$^a$ \\ 
SN\,2019tsf & 2020 Jan 21 & 82.9 & NOT+ALFOSC \\
SN\,2019tsf & 2020 Jan 26 & 87.0 & P60+SEDM \\
SN\,2019tsf & 2020 Feb 02 & 93.8 &  P60+SEDM\\
SN\,2019tsf & 2020 Feb 07 & 98.7 & P60+SEDM\\
SN\,2019tsf & 2020 Feb 15 & 106.4 & TNG+DOLORES \\
SN\,2019tsf & 2020 Feb 19 & 110.6 & Keck1+LRIS \\
SN\,2019tsf & 2020 Mar 22 & 141.9 & Keck1+LRIS \\
SN\,2019tsf & 2020 Apr 28 & 178.7 & NOT+ALFOSC \\
SN\,2019oys & 2019 Aug 29 & 0.96 & P60+SEDM \\
SN\,2019oys & 2020 Jan 27 & 150.0 & NOT+ALFOSC \\
SN\,2019oys & 2020 Feb 01 & 154.1 & P60+SEDM \\
SN\,2019oys & 2020 Feb 09 & 162.2 & P60+SEDM \\
SN\,2019oys & 2020 Feb 15 & 166.6 & NOT+ALFOSC \\
SN\,2019oys & 2020 Feb 19 & 172.0 & Keck1+LRIS \\
SN\,2019oys & 2020 Feb 24 & 176.8 & P60+SEDM \\
SN\,2019oys & 2020 Mar 22 & 203.5 & Keck1+LRIS \\
SN\,2019oys & 2020 Apr 15 & 227.0 & P60+SEDM \\
SN\,2019oys & 2020 May 01 & 243.3 & NOT+ALFOSC \\
SN\,2019oys & 2020 Aug 21 & 353.3 & Keck1+LRIS \\  \enddata
\tablenotetext{a}{This spectrum is from TNS provided by \citet{2019TNSCR2284....1M}.}
\end{deluxetable}

\clearpage

\onecolumn

\LTcapwidth=0.5\textwidth
\begin{longtable}{ccccc}
\caption{SN\,2019oys - radio observations\label{tab:Observations}}
\label{Table: radio data} 
\\ \hline 
\\
$\Delta t$ & Frequency & $F_{\nu}$ & Image RMS & Telescope\\ [1ex]
$[\textrm{Days}]$ & [\textrm{GHz}] & $[\textrm{mJy/beam}]$ & $[\textrm{mJy}]$ & \\ [1ex]
\hline 
\endfirsthead 
\hline 
$\Delta t$ & Frequency & $F_{\nu}$ & Image RMS & Telescope\\ [1ex]
$[\textrm{Days}]$ & [\textrm{GHz}] & $[\textrm{mJy/beam}]$ & $[\textrm{mJy}]$ & \\ [1ex]
\hline 
\endhead 
\hline
\multicolumn{3}{l}{\footnotesize Table \ref{Table: radio data} continues} \\
\hline
\endfoot
\hline
\endlastfoot	
$21.9$ & $15.5$ & $0.35 \pm 0.05$ & $0.04$ & AMI-LA \\  [0.5ex]
$25.7$ & $15.5$ & $0.37 \pm 0.05$ & $0.04$ & AMI-LA \\  [0.5ex]
$191$ & $15.5$ & $9.08 \pm 0.5$ & $0.06$ & AMI-LA \\  [0.5ex]
$198$ & $15.5$ & $10.0 \pm 0.5$ & $0.06$ & AMI-LA \\  [0.5ex]
$201$ & $23.5$ & $21.5 \pm 1.0$ & $0.05$ & VLA \\  [0.5ex]
$204$ & $15.5$ & $10.3 \pm 0.5$ & $0.05$ & AMI-LA \\  [0.5ex]
\hline
\end{longtable}

\twocolumn

\clearpage

\LTcapwidth=0.4\textwidth
\begin{deluxetable}{lcccc}
\tabletypesize{\scriptsize}
\tablecaption{Selected emission lines in SN~2019oys \label{tab:coronallines} }
\tablewidth{0pc}
\tablehead{
\colhead{Line ID} &\colhead{Wavelength}  &\colhead{Flux }                            & \colhead{FWHM} & \colhead{Comment}  \\
\colhead{}        &\colhead{(\AA)}   &\colhead{(10$^{-16}$ erg s$^{-1}$ cm$^{-2}$)} &\colhead{(\AA)}  &
}
    
\startdata
[Ne~{\sc v}]        &3346    & 3.2 		& 7.9 & Blended with [Ne~{\sc iii}] $ \lambda3342$ \\

[Ne~{\sc v}]		&3426  	 & 5.4 	& 4.7  & \\

$[$Fe~{\sc vii}]	&3586  	&   1.5 	& 5.2 	& \\

$[$O~{\sc ii}]		&3727  	&  	1.0 	 & 8.6 	& Weak \\

$[$Fe~{\sc vii}]	&3759  	&  1.8 	& 5.2 	& \\

$[$Ne~{\sc iii}]	&3869  	& 13.1	& 4.7	& Strong and narrow	\\

$[$He~{\sc I}]	& 3889  	& 6.1	& 12.2	& Broad\\

Ca~{\sc II}	&  3933 	& 8.8	& 17.9	& Ca II $\lambda$3933?, Broad \\

$[$Ne~{\sc iii}]	&3968   	& 10.7	& 9.1	&  \\

$[$Fe~{\sc v}]		&4072   	& 2.2 	& 7.3	& or more likely [S III] \\

H$\delta$		&4103  		& 9.7 	& 20.0 	&  Broad \\

H$\gamma$		&4340  	& 19.1 	& 21.2 	&  Broad	\\

$[$O~{\sc iii}]		&4363	 	& 4.9	& 6.2 	& \\

$[$Ar~{\sc xiv}]	&4412  	   	& 1.3  	& 5.8	& Confused with Fe II?\\

He~{\sc II}	& 4686 		& 1.3  	& 5.2	& \\

H$\beta$		&4861    	& 75.6	& 25.0	& Broad \\

$[$O~{\sc iii}]		&4959  	& 5.1	& 4.7	&\\

$[$O~{\sc iii}]		&5007 	& 16.8	& 5.0	&\\

$[$Fe~{\sc vii}]	&5158   & 3.1 	& 8.7	&\\

$[$Fe~{\sc vi}]		&5176  & 2.6	& 13 & Weak and blended, but present \\

[Ar~{\sc x}]	&5536  	& 1.3 & 10.7 & \\

[Fe~{\sc vii}] 	&5720  		& 1.6 	& 5.6 	&\\

$[$N~{\sc ii}]		&5755  	& 3.8	& 4.9	&  \\

He~{\sc i}    	&5876  	& 24.7	& 35.0    & Broad (+ Na I D) \\

$[$Fe~{\sc vii}] 	&6086  	& 2.4 	& 5.3 	&\\

$[$O~{\sc i}]		&6300  		& 1.9		& 6.5 &\\  

$[$S~{\sc iii}]		&6312  	&	1.3	& 12.1 & Present but weak	\\

$[$O~{\sc i}]		&6364  	& 	 1.3		& 10.1 & Somewhat blended \\

$[$Fe~{\sc x}]		&6375	& 3.3	& 6.1	&\\

H$\alpha$ &6563 & 737 & 34.7 & Broad and Strong\\

He~{\sc i}		&6678	& 3.3	& 24.1 & Broad, no narrow component\\

$[$S~{\sc ii}]		&6717   &	0.33&	5.9 &		 Weak \\

$[$S~{\sc ii}]		&6731   &	0.51 &    8.4   	& Weak\\

He~{\sc i} 	&7065		& 7.5	& 24.0   & Broad\\

$[$Ar~{\sc iii}] 	&7136  	& 0.76 	& 6.3	& Not in our model\\

$[$Fe~{\sc ii}]		&7155  	& 1.0		& 5.0 & Narrow, in model \\

$[$S~{\sc xii}]		&7611  		& 0.93 	& 5.8 	&\\

$[$Fe~{\sc xi}]		&7891  		& 2.1		& 5.3		&\\

O~{\sc i}		&8446   & 	14.7	& 37.6	& Broad, blended with Fe II and Pa \\

Ca~{\sc ii}		&8498,8542,8662  	& 5.8	& 33.7	& Redmost of the broad lines\\

$[$S~{\sc iii}]		&9069  	& 1.4	& 8.0	& \\

[Fe~{\sc ii}]	 &  8927 & 0.90 & 7.8 & Example of Fe II fluorescence   \\ 

[S~{\sc iii}]		& 9531 &  2.3 & 4.8 & \\

Paschen8 & 9546 & 12.2 & 38.2  &  \\

Paschen7 & 10036.7 & 14.2 & 53 &  \\
\enddata
\tablecomments{Fluxes are from the day 172 (+204) Keck spectrum, 
 absolute calibrated versus photometry and corrected for extinction in the Milky Way, see text.}
\end{deluxetable}

\end{document}